\documentstyle[11pt,newpasp,twoside,epsf]{article}
\reversemarginpar

\begin{document}

\title{Status and Prospects of Planetary Transit Searches:
\\	Hot Jupiters Galore}

\author{Keith Horne  {\it (kdh1@st-and.ac.uk) }  }

\affil{ Physics \& Astronomy,
         St.Andrews University,
         KY16 9SS, Scotland, UK}

\begin{abstract}  The first transiting extrasolar planet, orbiting
HD~209458, was a
Doppler wobble planet before its transits were discovered with a 10~cm
CCD camera.  Wide-angle CCD cameras, by monitoring in parallel the
light curves of tens of thousands of stars, should find hot Jupiter
transits much faster than the Doppler wobble method.  The discovery
rate could easily rise by a factor 10.  The sky holds perhaps 1000 hot
Jupiters transiting stars brighter than $V=13$. These are bright
enough for follow-up radial velocity studies to measure planet masses
to go along with the radii from the transit light curves.  I derive
scaling laws for the discovery potential of ground-based transit
searches, and use these to assess over two dozen planetary transit
surveys currently underway.  The main challenge lies in calibrating
small systematic errors that limit the accuracy of CCD photometry at
milli-magnitude levels.  Promising transit candidates have been reported by
several groups, and many more are sure to follow.
\end{abstract}

\section{Three Waves of Extrasolar Planet Discovery}

\subsection{Doppler Wobble Planets}

In the first wave of extrasolar planet discovery, Doppler wobble surveys have
sustained a discovery rate of 1 or 2 planets per month since the debut
of 51~Peg (Mayor \& Queloz 1995).  The extrasolar planet catalog now holds
over 100 Doppler wobble planets, filling in the top-left quadrant of
the planet mass $m$ vs. orbit size $a$ discovery space (Figure 1).  The
Doppler wobble planets have a roughly uniform distribution on the
$\log{m}$--$\log{a}$ plane, with two clear and interesting boundaries,
a maximum planet mass $m\la10~m_J$, and minimum orbit size
$a\ga0.03$~AU ($P>3$~d).  In their second decade, Doppler surveys will
extend the catalog to larger orbits and push down to somewhat lower
masses as precision radial velocities improve from 3 to 1~m~s$^{-1}$.


\begin{figure}[t]


\plotfiddle{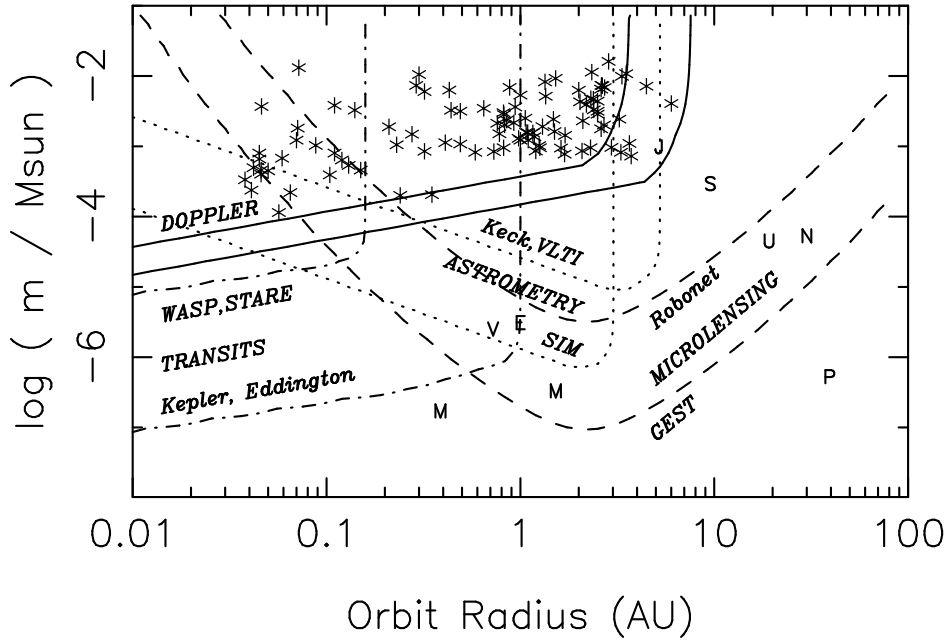}{8.1cm}{-90}{55}{55}{-200}{310}

\caption[] {\small
Planet detection thresholds on the planet mass vs. orbit radius plane
for four indirect methods including Doppler (-----), transits
($-\cdot-$) microlensing (- - -), and astrometry ($\cdot\cdot\cdot$).
The 9 solar system planets, and 100 extrasolar planets ($\ast$) are
also indicated.}
\end{figure}

\subsection{Hot Jupiter Transits}

A second wave, the hot Jupiter discovery era, is fast approaching.
The Doppler wobble surveys establish that hot Jupiters, with
$P\sim\ 4$~d and $a\sim\ 0.05$~AU, are orbiting around $\sim\ 1\%$ of
nearby sun-like stars.  A fraction of these, $P_t \approx R_\star/a
\sim\ 10$\%, will have orbital inclinations close enough to edge-on so
that the planet transits in front of its star.  We may therefore
expect that 1 star in 1000 should be ``winking'' at us.  During each
transit, the starlight dims by $\Delta f/f \approx
(r/R_\star)^2\sim\ 1\%$ for a time $\Delta t \approx (P R_\star/ \pi a)
\approx 3$ hours.  A CCD camera that can monitor $\sim\ 30,000$ stars in
parallel for $\sim\ 60$ clear nights should therefore be able to
discover $\sim\ 30$ hot Jupiters with periods out to $\sim\ 10$ days.

If ongoing transit surveys live up to these expectations, the hot
Jupiter discovery rate could soon rise to 10 or even 100 times that of
current Doppler wobble surveys.  Transits may reveal thousands of hot
Jupiters in the next 5 years.  A large catalog of hot Jupiters will
help to establish how their abundance, maximum mass, and minimum
period depend on star mass, age, metallicity, and environment.
Several hundred should be bright enough for radial velocity work to
establish masses to go along with radii from transits, establishing
relationships between mass, radius, orbit, and age.  Perhaps 10 will
be bright enough for atmospheric studies via scattered starlight
(Cameron et al. 1999), transit spectroscopy (Seager and Sasselov 2001;
Brown 2001; Charbonneau et al. 2002), and detection of infrared
thermal emission (Charbonneau 2003; Richardson et al. 2003).

\subsection{Transits from Space}

A third wave of discovery will arrive with the $\sim\ 2007$ launches of
Kepler (NASA) and Eddington (ESA).  These missions deploy CCD cameras
on wide-field space telescopes designed to detect transits of
Earth-sized planets with $\Delta f/f\sim\ 10^{-5}$.  Stellar variability
may be a limiting factor.  If successful, and depending on the
abundance of low-mass planets, the yield may be $\sim\ 10^4$ hot
Jupiters, $\sim\ 10^2$ ``hot Earths'', and a handful of ``habitable''
Earths.  The first discoveries will be hot Jupiters and Earths.  After
3--4 years, by $\sim\ 2011$, the catalog may include Earth-analogs in
the ``habitable'' zone.

\section{ The First Transits: HD~209458b }

Since hot Jupiters with $P\approx\ 4$~d have a 10\% transit probability,
we expect 1 in 10 of the shortest-period Doppler wobble planets to
exhibit transits.  Dramatic verification of this prediction came with
the discovery of the first extrasolar planet transits (Charbonneau et
al. 2000).  HD~209458 dims by 1.6\% for 3 hours every 3.5 days.  This
star is so bright ($V=7.8$) that the transits could be discovered with
a remarkably small (10~cm!)  wide-angle ($6^\circ$) CCD camera
(STARE).  The wide field is essential for high-precision differential
photometry of stars this bright, so that comparably bright comparison
stars can be measured simultaneously.

The HD~209458 transits were quickly confirmed by several groups (Henry
et al.\ 2000; Jha et al.\ 2000).  The most spectacular light curves by
far were captured by using the HST/STIS spectrograph with a wide slit
(Brown et al.\ 2001).  The transit shape, recorded in exquisite
detail, fits an immaculate limb-darkened star occulted by the circular
silhouette of the planet, yielding the orbit inclination
$i=86^\circ.6\pm0^\circ.2$ and planet radius $r=1.35\pm0.06~r_J$.
With an rms accuracy better than $10^{-4}$, significant limits were
placed on moons ($r_{\rm moon}<1.2~r_\oplus$) and rings ($r_{\rm
ring}<1.8~r_\oplus$).  With $m\sin{i}=0.69~m_J$ from the star's
Doppler wobble, this hot Jupiter is clearly a ``bloated'' gas giant.
Transit spectroscopy (Seager and Sasselov 2001; Brown 2001) detects
Na~I absorption from the extrasolar planet atmosphere (Charbonneau et
al. 2002).

\section{Scaling Laws for Planetary Transit Surveys}

Discovery of transits depends on the planet (mass $m$, radius $r$,
orbit size $a$), on the star (mass $M_\star$, radius $R_\star$,
luminosity $L_\star$, distance $d$, galactic latitude $b$, dust
extinction $K$) and on experimental parameters (aperture $D$, field of
view $\theta$, quantum efficiency $Q$, bandwidth $\Delta\lambda$,
angular resolution $\Delta\theta$, sky brightness $\mu_{\rm sky}$,
duration $t$, duty cycle $f$).  Scaling laws can help in optimizing
and comparing the discovery potentials of current experiments.

The planet catch ($N_p$ planets) is
\begin{equation}
\frac{dN_p}{d\log{a}\ d\log{m}}
	\approx \left( \frac{\theta^2 d^3}{3} \right)
	\left( \frac{n_\star}{ e^{d\left|\sin{b}\right|/H} }\right)
	\left( \frac{df_p}{d\log{a}\ d\log{m}}\right)
	\left( \frac{R_\star}{a} \right)
\ .
\end{equation}
The four factors are:
(1) The survey volume, $\left(\theta^2 d^3/3\right)$,
covering a solid angle $\theta^2$ out to distance $d$.
(2) The star number density
$ n_\star\ e^{-d\left|\sin{b}\right|/H}$,
with a local density
$n_\star \sim\ 0.02$~F,G,K~stars~pc$^{-3}~M_\odot^{-1}$,
and galactic disk scale height $H\approx300$~pc.
(3) The number of planets per star,
$\left(df_p/d\log{a}\ d\log{m}\right)\equiv\eta_p\approx0.05$,
for $a>0.03$~AU ($P>3$d) and $m<10~m_J$.
This gives 0.007 hot Jupiters (3--5d, 1--10~$m_J$) per star,
consistent with the findings of Doppler wobble surveys.
(4) The orbit alignment probability, $P_t \approx\ R_\star/a$.

The signal-to-noise ratio for transit detection is
\begin{equation}
\frac{S}{N} \approx \frac{ (r/R_\star)^2\ f_\star }
	{ \left( f_\star + f_{\rm sky} \right)^{1/2} }
	\left( \frac{P\ R_\star}{\pi\ a } \frac{t\ f}{ P } \right)^{1/2}
\ .
\end{equation}
The transit depth $\Delta f \approx (r/R_\star)^2 f_\star$
must be detected against Poisson noise
from the star and sky photons.
The time available to do this is the transit duration
$\Delta t \approx ( P R_\star / \pi a )$,
and the $S/N$ improves as $\sqrt{\Delta t\ N_t}$, where
$N_t \approx t f / P$ is the number of transits observed in time $t$.
The number of sky and star photons scale as
$f_{\rm sky} \propto \left( D^2\ Q\ \Delta\lambda \right)
		\left(  \Delta\theta^2\ \mu_{\rm sky} \right)$
and $f_\star \propto \left( D^2\ Q\ \Delta\lambda \right)
		\left( L_\star d^{-2} e^{-Kd} \right)$,
respectively.

The survey distance $d$ is the maximum
at which transits are detectable (e.g., $S/N>10$).
In ground-based surveys, sky noise sets this
faint limit, and the planet catch then scales as
\begin{equation}
\frac{dN_p}{d\log{a}\ d\log{m}}
\propto
	\left( \frac{\eta_p\ r^3}{ a^{7/4}} \right)
	\left( \frac{ L_\star^{3/2}\  n_\star }
		{ R_\star^{5/4}\  e^{3Kd/2}  } \right)
	\theta^2
	\left( \frac{ D^2 Q\ \Delta\lambda\ f\ t }
		{ \Delta\theta^2\ (S/N)^2 \mu_{\rm sky} }\right)^{3/4}
\ ,
\end{equation}
where the planet, star, and experimental parameters are grouped.

To estimate the discovery rate,
consider a typical hot Jupiter
($r=r_J$, $P=4$~d, $a=0.05$~AU)
orbiting a sun-like star
($L_\star=L_\odot$, $M_\star=M_\odot$, $R_\star=R_\odot$).
For a fiducial set of transit survey parameters
(
$D=1$m,
$\theta=1^\circ$,
$\Delta\theta=1$~arcsec,
$Q=0.3$,
$\Delta\lambda=2000$\AA,
$t=60$~d,
$f=0.25$,
$S/N=10$,
$\mu_{\rm sky}=18$~mag~arcsec$^{-2}$,
$K=0.5$~mag~pc$^{-1}$),
evaluation of the above expressions
indicates that hot Jupiter transits
can be detected on sun-like stars down to $V \approx 18$,
out to $d \approx 2.5$~kpc, at a rate of 12 planets per month.


\begin{figure}[t]

\plotfiddle{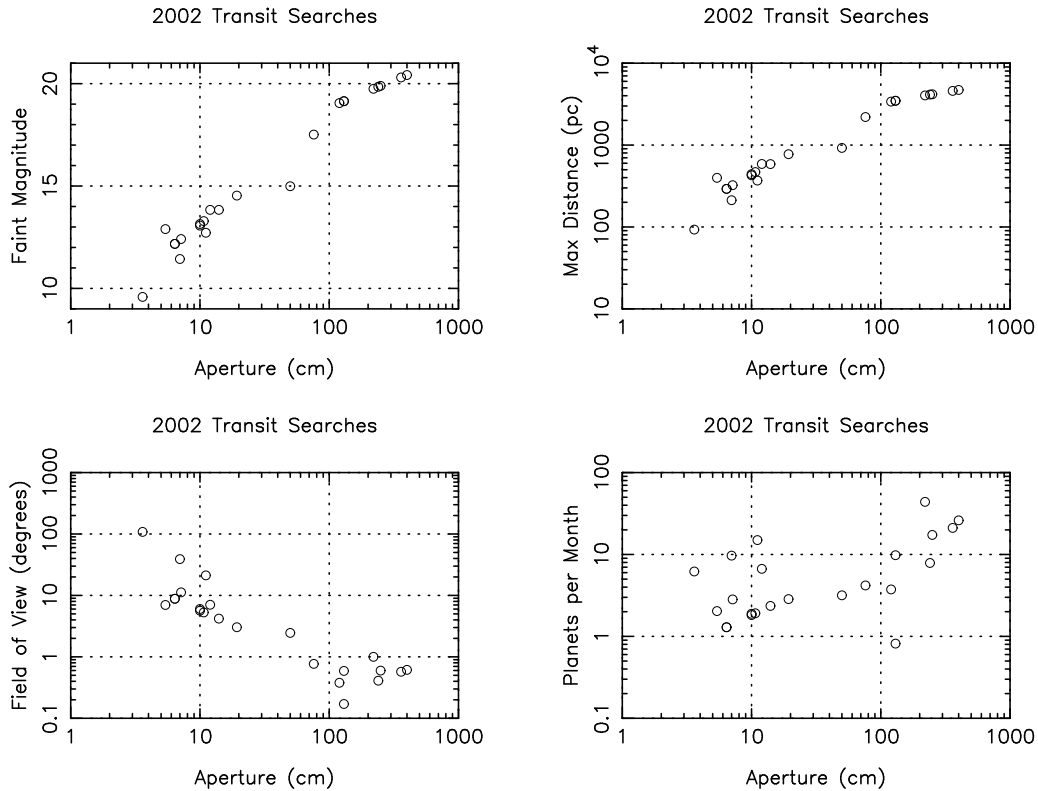}{9.5cm}{-90}{55}{55}{-200}{310}

\caption[] {\small
Parameters and discovery potential for the 2002 transit surveys
listed in Table 1.
}
\label{fig:two}
\end{figure}

\section{Transit Searches Wide and Deep}

For this review, two dozen teams returned e-mail questionaires
providing information on their transit search experiments.  The large
number of teams makes it impossible to report more than summary
information, which is collected in Table~1, and summarized in Figure~2.

The experiments break into two main groups, wide and deep.  Because
the planet catch scales as $\theta^2D^{3/2}$, wider fields of view
enable small telescopes to compete.  For each experiment I have used
the scaling laws to evaluate the faint magnitude, the maximum
distance, and the hot Jupiter discovery rate.  Summing over all
experiments, the potential discovery rate approaches $\sim\ 200$ planets
per month.

The wide-angle survey teams follow STARE and Vulcan in deploying small
($\sim\ 10$~cm) wide-angle ($\sim\ 10^\circ$) CCD cameras.  These
experiments use a CCD pixel size $>1$~arcsec, sacrificing angular
resolution to expand the field of view.  The faint limit at 
$V \sim\ 12$--13
reaches to $d \sim\ 300$--500~pc, 
comparable to the galactic disk scale height, 
so that target fields cover the entire sky.

The deep survey teams employ existing CCD cameras, often mosaics, on
established (1--4~m) telescopes.  The faint limits at $V\sim\ 19$--21 
reach to $d\sim\ 4$--5~kpc (limited by dust), so that galactic plane and 
open 
cluster fields are primary targets.

Our estimates suggest typical discovery rates of 3--10 planets per
month for both wide and deep surveys.  For multi-CCD mosaic cameras on
2--4~m telescopes, a discovery rate of 30 planets per month would be
possible, if these telescopes could be dedicated to transit searches.
It is delightful to realize that 10~cm telescopes can compete with a
4~m in the discovery of hot Jupiters.

\begin{table}[t]

\begin{center}
{\bf Table 1. 2002 Planetary Transit Surveys}
\\[2mm]
\begin{tabular}{@{\extracolsep{-2mm}}rlrrrlrrrrrr}
\hline
    & program & $D$ & $F/$ & $\theta$
            & CCD & pixel
  & sky & star & $d$ &  stars & planets \\
  & & cm &  & deg & kxk 
& arcsec & mag & mag & pc &  $10^3$ & /month \\[1mm]
\hline
\hline
     $  1$ &   PASS       & $   3.6$ & $  1.4$ & $ 108$ & 1x1$\times$15
   & $ 98$ & $  5.6$ & $  9.6$ & $    93$ & $  18$ & $     6$ \\
     $  2$ &   Vulcan-S   & $   5.4$ & $  5.6$ & $   6.98$ & 4x4
   & $  6.1$ & $ 11.7$ & $ 12.9$ & $   397$ & $  6$ & $     2$ \\
     $  3$ &   HAT-1      & $   6.4$ & $  2.8$ & $   8.84$ & 2x2
   & $ 15.5$ & $  9.6$ & $ 12.2$ & $   292$ & $  4$ & $     1$ \\
     $  4$ &   WASP0      & $   6.4$ & $  2.8$ & $   8.84$ & 2x2
   & $ 15.5$ & $  9.6$ & $ 12.2$ & $   292$ & $  4$ & $     1$ \\
     $  5$ &   ASAS-3     & $   7.1$ & $  2.8$ & $  11.2$ & 2x2$\times2$
   & $ 13.9$ & $  9.9$ & $ 12.4$ & $   323$ & $  8$ & $     3$ \\
     $  6$ &   PPS        & $  10.0$ & $  2.8$ & $   5.66$ & 2x2
   & $  9.9$ & $ 10.6$ & $ 13.1$ & $   441$ & $  5$ & $     2$ \\
     $  7$ &   PSST       & $  10.7$ & $  2.8$ & $   5.29$ & 2x1
   & $  9.3$ & $ 10.8$ & $ 13.3$ & $   468$ & $  5$ & $     2$ \\
     $  8$ &   STARE      & $  10.0$ & $  2.9$ & $   6.03$ & 2x2
   & $ 10.7$ & $ 10.5$ & $ 13.1$ & $   427$ & $  5$ & $     2$ \\
     $  9$ &   SuperWASP  & $  11.1$ & $  1.8$ & $  21.2$ & 2x2$\times$5
   & $ 16.7$ & $  9.5$ & $ 12.7$ & $   368$ & $ 43$ & $    15$ \\
     $ 10$ &   Vulcan     & $  12.0$ & $  2.5$ & $   7.04$ & 4x4
   & $  6.2$ & $ 11.6$ & $ 13.8$ & $   587$ & $  19$ & $     7$ \\
     $ 11$ &   RAPTOR     & $   7.0$ & $  1.2$ & $  39.1$ & 2x2$\times$4
   & $ 34.4$ & $  7.9$ & $ 11.4$ & $   212$ & $ 28$ & $    10$ \\
     $ 12$ &   RAPTOR-F   & $  14.0$ & $  2.8$ & $   4.19$ & 2x2
   & $  7.4$ & $ 11.3$ & $ 13.8$ & $   586$ & $  7$ & $     2$ \\
     $ 13$ &   BEST       & $  19.3$ & $  2.7$ & $   3.04$ & 2x2
   & $  5.3$ & $ 12.0$ & $ 14.5$ & $   774$ & $  8$ & $     3$ \\
     $ 14$ &   SSO/APT    & $  50$ & $  1.0$ & $   2.46$ & 0.8x1.1
   & $  9.4$ & $ 10.7$ & $ 15.0$ & $   923$ & $  9$ & $     3$ \\
\hline
     $ 15$ &   TeMPEST    & $  76$ & $  3.0$ & $   0.77$ & 2x2
   & $  1.35$ & $ 15.0$ & $ 17.5$ & $  2200$ & $  12$ & $     4$ \\
     $ 16$ &   PISCES     & $ 120$ & $  7.7$ & $   0.38$ & 2x2$\times4$
   & $  0.33$ & $ 17.1$ & $ 19.1$ & $  3395$ & $  11$ & $     4$ \\
     $ 17$ &   ASP        & $ 130$ & $ 13.5$ & $   0.17$ & 2x2
   & $  0.30$ & $ 17.1$ & $ 19.1$ & $  3477$ & $  2$ & $     1$ \\
     $ 18$ &   OGLE-III   & $ 130$ & $  9.2$ & $   0.59$ & 2x4$\times8$
   & $  0.26$ & $ 17.1$ & $ 19.1$ & $  3477$ & $  28$ & $    10$ \\
     $ 19$ &   GOCATS     & $ 220$ & $  ?$ & $   1.00$ & 4x4$\times4$
   & $  0.44$ & $ 17.1$ & $ 19.7$ & $  4036$ & $  126$ & $    44$ \\
     $ 20$ &   STEPSS     & $ 240$ & $  ?$ & $   0.41$ & 4x2$\times8$
   & $  0.18$ & $ 17.1$ & $ 19.8$ & $  4131$ & $  23$ & $     8$ \\
     $ 21$ &   UStAPS    & $ 250$ & $  3.0$ & $   0.60$ & 2x4$\times4$
   & $  0.37$ & $ 17.1$ & $ 19.9$ & $  4176$ & $  50$ & $    17$ \\
     $ 22$ & EXPLORE-S  & $ 400$ & $  2.9$ & $   0.61$ & 2x4$\times8$
   & $  0.27$ & $ 17.1$ & $ 20.4$ & $  4707$ & $  75$ & $    26$ \\
     $ 23$ & EXPLORE-N  & $ 360$ & $  4.2$ & $   0.57$ & 2x4$\times12$
   & $  0.21$ & $ 17.1$ & $ 20.3$ & $  4586$ & $  61$ & $    21$ \\
\hline \\
\multicolumn{11}{l}{Total planets per month:} & 191
\end{tabular}
\end{center}
(See \verb"http://star-www.st-and.ac.uk/~kdh1/transits/table.html"
for additional information and links to project web pages.)
\end{table}

The discovery rates in Table~1 may be good to a factor of 2,
and should be regarded as ideal performance benchmarks for dedicated 
transit surveys, 
indicating also the relative discovery potential of different experiments.
Actual performance will be degraded by many effects
(crowding, moonlight, airmass, seeing, vignetting,
weather, sub-optimal data analysis, competing observing programs).
As a sanity check,
Table~1 estimates 19,000 stars with $V<13.8$
and 7 planets per month for the Vulcan experiment.
For comparison, Borucki et al. (2001) analyzed
6000 stars down to $V=13$, finding over 100 variables,
about 50 eclipsing binaries, and 3 planetary transit candidates
(all rejected by follow-up spectroscopy).
Why this difference between ideal and actual performance?
From Figure~6 in Borucki et al. 2001,
Vulcan light curves achieved 1.5 times the photon noise
for $V_{\rm sky}=11.1$.
The sky is 1.6 times brighter than $V_{\rm sky} =11.6$ from Table~1,
probably because Vulcan's PSF is wider than
the 2-pixel (12 arcsec) FWHM gaussian adopted here.
This should cut the survey volume by
$(1.5)^{3/2}(1.6)^{3/4} = 2.6$.
In fact moving the faint limit from $V=13.8$ to $V=13.0$ 
cuts the volume by a factor 3.
This explains the difference between 19,000 and 6000 stars,
and validates our star count estimates.
We still expect $7/3\sim 2.3$ planets per month.
We adopt $S/N>10$ for transit detection,
and the planet catch scales as $(S/N)^{-3/2}$.
Perhaps Vulcan's effective threshold is even higher
given that the elaborate self-calibration used to
reduce systematic errors also suppresses weak transit signals.
This comparison illustrates both the difficulty of
optimizing performance, and the benefits of doing so,
since most discoveries will be near the faint limit
of the survey.

\subsection{Special Target Transit Surveys}

Three teams report using a strategy targeting specific stars to
enhance their chances of planet discovery.  The TEP team targets
low-mass eclipsing binaries, enhancing the transit probability if the
binary and planetary orbits are co-aligned, and the sensitivity to
small planets due to the small stellar radii.  CM~Dra is now
thoroughly probed for circumbinary planets down to $\sim\ 3~r_\oplus$
(e.g., Deeg et al 1997; Doyle et al. 2000), and several other systems
are under study.

Greg Henry (TSU) uses several robotic photometric telescopes at
Fairborn Observatory to follow-up Doppler wobble planets in search of
transits at known times of conjunction.  This has lead for example to
independent discovery of the transits of HD~209458b (Henry et
al. 2000).

Tim Castellano and Greg McLaughlin (transitsearch.org) are
coordinating a network of amateur observers to target stars with known
planets.  The times of transit, and the probability of a planet
transit for a given star, are known in advance, thus limiting the
observation time and data analysis.

\subsection{ Globular Cluster Transit Surveys }

HST can resolve main-sequence stars in the crowded cores of globular
clusters.  Staring for 8 days at 47~Tuc, HST monitoring of 34,000
$V=17$--21.5 main-sequence stars should have found 17 transits, but in
fact found none (Gilliland et al.\ 2000).  Hot Jupiter formation
and/or survival is evidently inhibited, perhaps by low metallicity
(Gonzalez 1998; Gonzalez et al.\ 2001; Santos, Israelian \& Mayor 2001),
ultraviolet evaporation (Armitage 2000), or collisional disruption
(Bonnell et al.\ 2001) of the proto-planetary disks in this crowded
stellar environment.

\subsection{Open Cluster Transit Surveys}

Janes~(1996) recommended open clusters as ideal targets for planet
transit surveys.  Teams currently using 1--2.5~m telescopes to hunt
transits in open cluster fields include PISCES (Whipple 1.2~m,
Mochejska et al. 2002), GOCATS (LPL 1.6/2.2~m, C. Barnes), STEPSS (MDM
2.4~m, J. Burke et al.)  and UStAPS (INT 2.5~m, Street et al. 2002).  The
clusters (number of nights) reported to be under analysis are PISCES:
N6791(25), GOCATS: M35+M67(15), STEPSS: N1245(19), UStAPS:
N6819(20/3), N6940(20/3), N7789(20/3+10).  Although field stars
usually dominate these surveys, the clusters provide samples of stars
with a common age, metallicity and distance.

\subsection{Galactic Plane Transit Surveys}

Large (1--4~m) telescopes equipped with wide-field (0.5--1$^\circ$) CCD
cameras are capable probes of planetary transits of stars at large
distances (2--4~Kpc).  The galactic plane provides a high density of
stars in the long narrow survey volume.  These deep surveys will
measure the abundance and period distribution of hot Jupiters in a
variety of stellar populations throughout the galaxy well beyond the
solar neighborhood.

The OGLE III team, using their 1.3~m microlensing survey telescope's
new $0.6^\circ$ CCD camera, have monitored $\sim\ 52,000$ galactic disk
stars for 32 nights, and report 59 transit candidates with periods
ranging from 1 to 9 days (Udalski et al. 2002a,2002b).  The EXPLORE team,
using the CTIO 4~m and CFHT 3.6~m, have observed two galactic plane
fields, finding 3 possible planetary transit candidates
(Mallen-Ornelas et al. 2002).

Among the candidates already identified may be the first hot Jupiters
discovered from their transit signatures.  Since Jupiters, late M
dwarfs, and brown dwarfs have similar radii, and partial eclipses can
mimic transits by smaller bodies, confirmation of these extrasolar
planet candidates now awaits radial velocity follow-up with 
$\sim\ 1~$km~s$^{-1}$ precision to detect or rule out the star wobble
signature of stellar and brown dwarf companions.

\subsection{Wide-Angle Transit Surveys}

Wide-angle transit surveys, following in the footsteps of Vulcan and
STARE, may offer the most exciting discovery potential because they
target bright stars for which follow-up radial velocity work can
measure masses to go along with the radii from transits.  The main
challenge is to achieve $\sim\ 10^{-3}$mag accuracy in differential
photometry over a very wide field of view, in which airmass, transparency,
differential refraction, seeing, and even the heliocentric time
correction, all varying significantly across the field.

If the requisite accuracy can be achieved (e.g., Borucki et al.\ 2001),
these surveys should discover hot Jupiters transiting thousands of
bright nearby main-sequence F, G, K, and M stars.  Given the modest
survey depth ($\sim\ 400$~pc), the targets are solar neighborhood stars
distributed over the entire sky.  For a conservative estimate, assume
that $V=7.8$ HD~208458b is the brightest.  Since each magnitude
quadruples the number of stars in the survey volume, if there is 1
brighter than $V=8$, there should be 4 down to $V=9$, 16 to $V=10$,
64 to $V=11$, 250 to $V=12$, and 1000 to $V=13$.

How long might it take to find the 1000 brightest stars transited by
  hot Jupiters?  There should be 2--3 in each $10^\circ\times10^\circ$
field.  Assuming 2 months per field, it would take over 60 years for a
single $10^\circ$ CCD camera to survey the entire sky.
Fortunately, with a dozen experiments already underway (Table 1), the
hot Jupiter discovery era will likely be complete to $V = 13$ in
$\sim\ 5$~years.

\subsection{When will the Fun Begin?}

Expectations are high, but the number of new planets
revealed by transits is still zero.
Is something amiss?
Many teams are working hard (Table~1),
but the data analysis and computer processing requirements
for a dedicated transit survey are challenging and
have not yet been achieved by most of these groups.
Several teams have reported their first batches
of transit {\em candidates}, and it is
likely that the first new planets are in these lists.
However, we must expect that many (most?) of the
transit candidates will be false alarms or transit mimics 
(stellar variability, parital eclipses,
brown dwarf or white dwarf eclipses).
Weeding out the mimics requires follow-up observations, e.g.
multi-colour eclipse photometry, low-resolution spectroscopy,
and medium-precision ($\la\ 1$~km~s$^{-1}$) radial velocities,
which are already underway.

What if discovery rates fail to meet our expectations?
We must allow, of course, for the reduced efficiency
of actual observing programs,
and for some losses due to necessary shortcuts in the data analysis,
but if discovery rates are still too low
this will also be an interesting result.
The hot Jupiter abundance found in Doppler wobble surveys
may be higher than in some of the deeper fields
targeted in transit surveys.
It is also possible that HD~209458b is atypically large,
and that most of the hot Jupiters are smaller and hence
harder to detect by means of transits.
Whatever the outcome, we should not have to wait long.
Discoveries will materialize or not within a year.

%

\section{Conclusion}

With Doppler wobble surveys now reaching for long-period planets, is
the extrasolar planet field approaching a discovery plateau?
Emphatically no.  The era of hot Jupiters is about to open with
discovery rates that should rise to 10 or 100 times the 1 or 2
planets per month from ongoing Doppler wobble surveys.
The next 5 years should bring us hot Jupiters galore.

\section*{Update:}
Spectroscopic follow-up (Konacki et al. 2003) confirms that
OGLE-TR-56b is an $0.9~m_J$ planet with a 1.2~d period, making this
the first exo-planet to be discovered by means of its transits.
The OGLE team have also reported 62 additional candidates
(Udalski et al. 2002c)
in the Carina region of the Galactic disk ($\ell \approx 290^\circ$).


\end{document}